\documentclass[doublecol]{epl2}
\usepackage{amssymb,amsmath,amscd}
\usepackage{mathrsfs}
\usepackage{stmaryrd,esint}
\usepackage{subfig}
\usepackage{ntheorem}
\usepackage{tabularx} 
\usepackage{calc} 
\usepackage{graphicx}

\def\bf{\mathbf}
\def\bfnabla{\mbox{\boldmath $\nabla$}}

\newcommand\Rm{{{\rm Rm}}}

\renewcommand{\revision}[1]{#1}

\title{Toward an asymptotic behaviour\\ of the ABC dynamo}
\shorttitle{Toward an asymptotic behaviour of the ABC dynamo}
\author{Isma\"el Bouya\inst{1,2} \and Emmanuel Dormy\inst{3}}

\institute{                    
  \inst{1} KTH Mechanics, Osquars Backe 18, SE-100 44 Stockholm, Sweden.\\
  \inst{2} UMR 7586 Institut de math\'ematiques de Jussieu, Analyse fonctionnelle,
    Universit\'e Paris Diderot, 2 Avenue de France, 75013 Paris, France.\\
  \inst{3} MAG (ENS/IPGP), LRA, D\'epartement de Physique, Ecole Normale Sup\'erieure, 24,
    rue Lhomond, 75231 Paris Cedex 05, France.
}

\abstract{
The ABC flow was originally introduced by Arnol'd to investigate Lagrangian
chaos. It soon became the prototype example to illustrate magnetic field amplification via fast dynamo
action, i.e. dynamo action exhibiting magnetic field amplification on 
a typical timescale independent of the electrical resistivity of the medium.
Even though this flow is the most classical example for this important class
of dynamos (with application to large scale astrophysical objects), it was
recently pointed out \cite{bouya2013revisiting} that the fast dynamo nature
of this flow was unclear, as the growth rate still depended on the magnetic
Reynold number at the largest values available so far (\(
\Rm=25000\)). Using state of the art high performance computing, we present
high resolution simulations (up to \(4096^3\)) and extend the value of
\(\Rm\) up to  \( 5\cdot10^5\).
Interestingly, even at these huge values, the growth rate of the leading eigenmode
still depends on the controlling parameter and an asymptotic regime is not
reached yet. We show that the maximum growth rate is a decreasing function
of \(\Rm\) for the largest values of \(\Rm\) we could achieve (as
anticipated in the above-mentioned paper). 
Slowly damped oscillations might indicate either a new mode crossing
or that the system is approaching the limit of an essential spectrum.
}

\begin{document}

\maketitle

\section{Introduction}
Fifty years after the ABC flow has been introduced in the seminal work of
Arnol'd \cite{arnold1965topologie}, as a prototype for Lagrangian chaos, its
properties as a fast dynamo are still unclear. In a recent study
\cite{bouya2013revisiting}, we stressed that contrary to earlier
expectations, this flow still does not act as a fast dynamo for \(\Rm
\simeq 25000\). The same year \cite{jones2013dynamo}, introduced a detailed
study of the symmetries of the various dynamo branches up to \(\Rm=10^4\).
Here, we investigate the kinematic dynamo action associated with the 
ABC-flow up to \(\Rm=5\cdot 10^5\). Such extreme values require very high
spectral resolutions (up to \(4096 ^3\) modes) and state of the art
parallel computing.

\begin{figure*}[t]
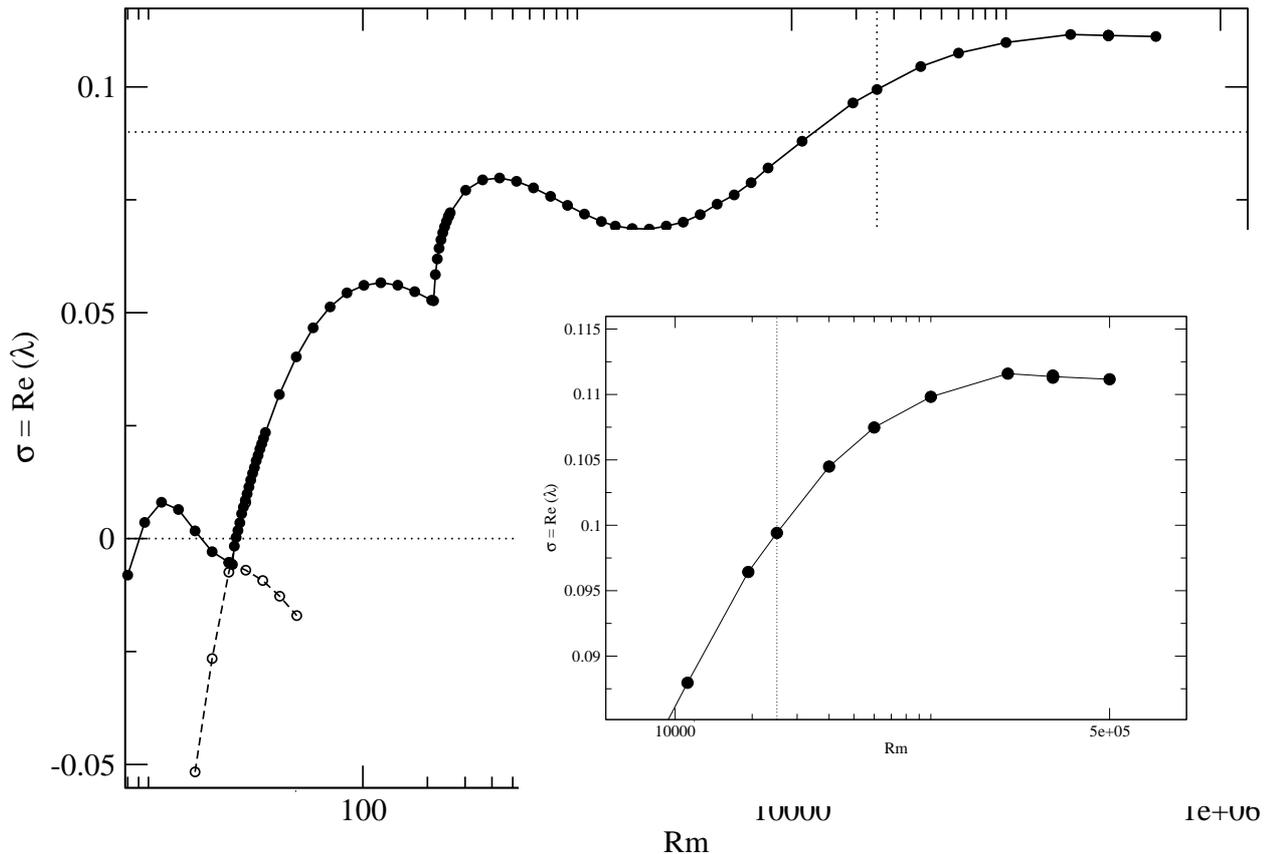

  \includegraphics[width=165mm]{abc_Rm_completee_plus}
  \begin{picture}(100,0)
    \put(200,50){\includegraphics[width=85mm]{abc_Rm_completee_zoom}}
  \end{picture}
  \caption{\label{fig:extension} Plot of the magnetic field growth rate 
  as a function of \(\Rm\), up to $\Rm = 5 \cdot 10 ^5$. The
  inset presents a closer view on the range $2 \cdot 10^5$ -- $2 \cdot 10^6$, which
  stresses the decrease of the growth rate at our larger values of \(\Rm\).
\revision{The horizontal dotted line at $\sigma \simeq 0.09$ corresponds to the theoretical 
upper bound provided by the topological entropy ($h_{\rm line}$).}
}
\end{figure*}

\section{Governing equations}
The time evolution of the magnetic field in a conducting medium (such as
an ionized astrophysical plasma) is governed by the magnetohydrodynamics
equations.
If one assumes that the magnetic field is weak enough not to influence the
fluid flow, a single equation, known as the induction equation, governs the
time evolution of the 
solenoidal magnetic field under a prescribed flow 
\begin{equation}
  \frac{\partial \bf B}{\partial t} = \bfnabla \times \left( \bf u \times \bf B
    - \Rm^{-1}\, \bfnabla \times \bf B \right) \, .
\label{induct}
\end{equation}
Finding exponentially growing solutions to this equation is known as the
kinematic dynamo problem. 
We consider here the ABC-flow
(\cite{arnold1965topologie,henon1966topologie}), which takes the form
\begin{align}
  \bf u  = & \,
  (A \sin z + C \cos y) \, \bf e_x    \\
  +&\, (B \sin x + A \cos z) \, \bf e_y\nonumber \\
  +&\, (C \sin y + B \cos x) \, \bf e_z \nonumber \, .
\end{align} 
We want to assess its fast dynamo property, i.e. the independence of the
growthrate on \(\Rm\) in the limit \(\Rm\rightarrow\infty\).
We restrict our attention to configurations in which the magnetic field has the
same periodicity as the flow (i.e. \(2\pi\)-periodic in all directions of
space, see \cite{archontis2003numerical} for extensions) and the weights of
the three symmetric Beltrami components are of equal strength
(\(A=B=C\equiv 1\)).

The simulations presented in this paper were performed using a modified
version of a code originally developed by
\cite{galloway1984numerical}. It uses a fully spectral method with
explicit mode coupling, which we parallelized using domain decomposition in the
spectral space (see \cite{bouya2013revisiting}).

\section{Numerical simulations up to \(\Rm= 5 \cdot 10^5\)}

In order to try to approach an asymptotic behaviour, we extend our previous
study of the variation of the fastest growth rate as a 
function of the magnetic Reynolds number \cite{bouya2013revisiting} up to
\(\Rm=5\cdot 10^5 \). 
Each simulation involves \(N^3\) Fourier modes. Simulations up to \(\Rm=6\cdot 10^4\)
were performed with resolutions \(N=512\) 
and \(N=1024\) in order to check convergence. Simulations up to
\(\Rm=3 \cdot 10^5\) were performed with resolutions \(N=1024\)
and \(N=2048\) and the highest \(\Rm\) we were able to perform, 
\(\Rm=5 \cdot 10^5\) was validated using \(N=2048\)
and \(N=4096\). 
\revision{All simulations were initialized with a random divergence
  free initial seed field, with the exception our our largest and most
  expensive simulation, \(\Rm=5\cdot 10^5\), which was started using
  the final stage of \(\Rm=3\cdot 10^5\).}
The resulting plot of the fastest growth rate is displayed in Figure~\ref{fig:extension}.

Our former study, up to \(\Rm = 25000\) revealed a growth rate \(\sigma
\simeq 0.1\) for the magnetic energy. This value is in excess of 
the theoretical upper bound provided by the so-called topological entropy \revision{($h_{\rm line}\simeq 0.09$)
\cite{klapper1995rigorous,childress1995stretch}.}
We therefore anticipated the necessity of a decrease of the growth rate at
larger (not yet available) values of \(\Rm\).

Enlarging the graph of the evolution of the growth rate for large \(\Rm\)
(see the inset in Figure~\ref{fig:extension}) clearly highlights
that the maximum growth rate, indeed reaches a local maximum
around \(\Rm\simeq 2 \cdot 10^5\), and then decreases with \(\Rm\) above this value. 
The imprecision on the growth rate is associated with slowly
damped oscillations, which are present at large magnetic Reynold
number (see below).

It is striking to note that even for \(\Rm = 5\cdot 10^5\), the growth rate
has not settled to an asymptotic value. Not only does it still vary with the
controlling parameter \(\Rm\), but it is also still significantly larger
(\(\sigma\simeq 0.11\)) than the theoretical upper-bound
(\(h_{\rm line} \simeq 0.09\)).

\section{Cross sections of the $(x-y)$ plane with rescaled coordinates}

\begin{figure}
  \onefigure[width=80mm]{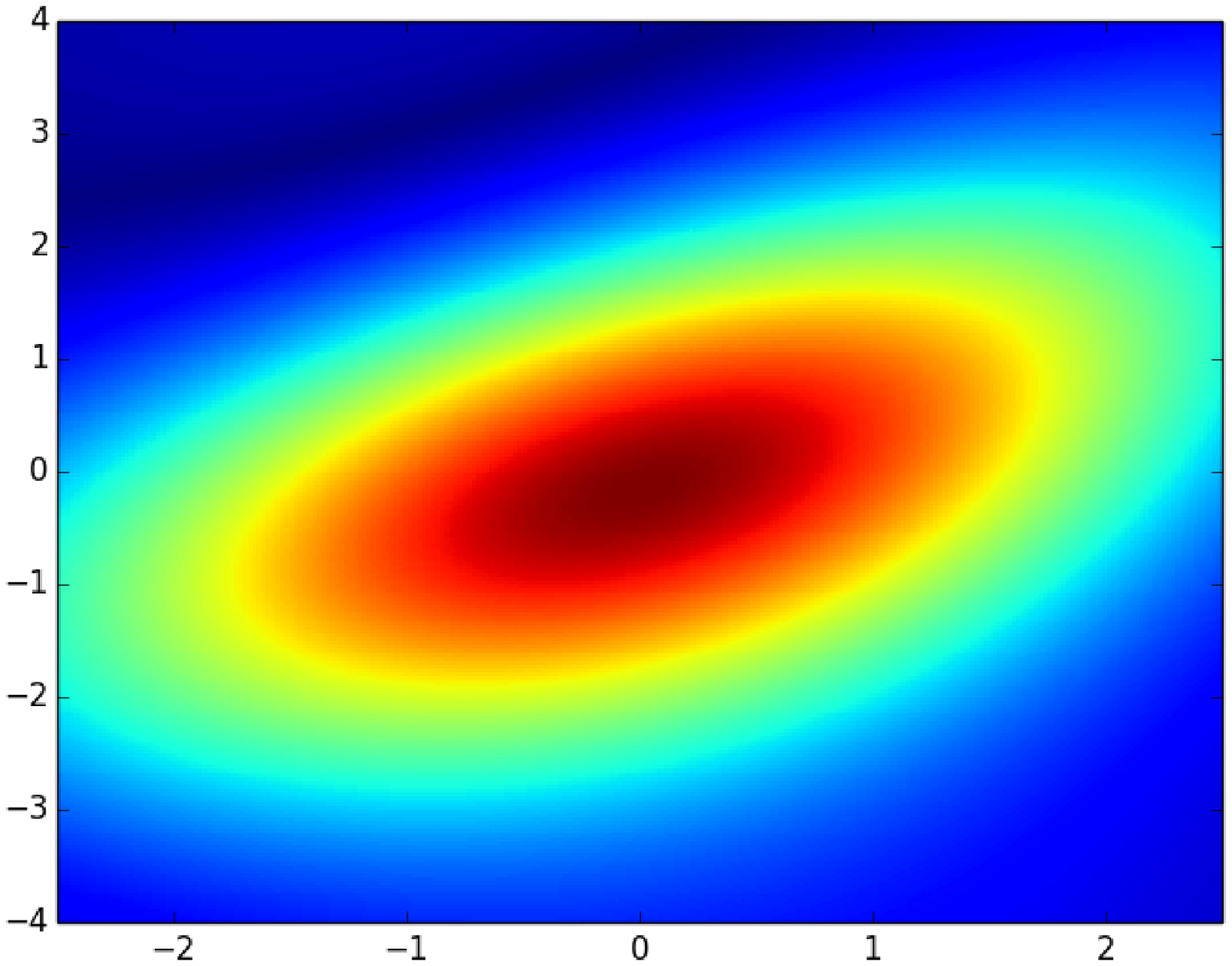}\vskip -3mm
  \onefigure[width=80mm]{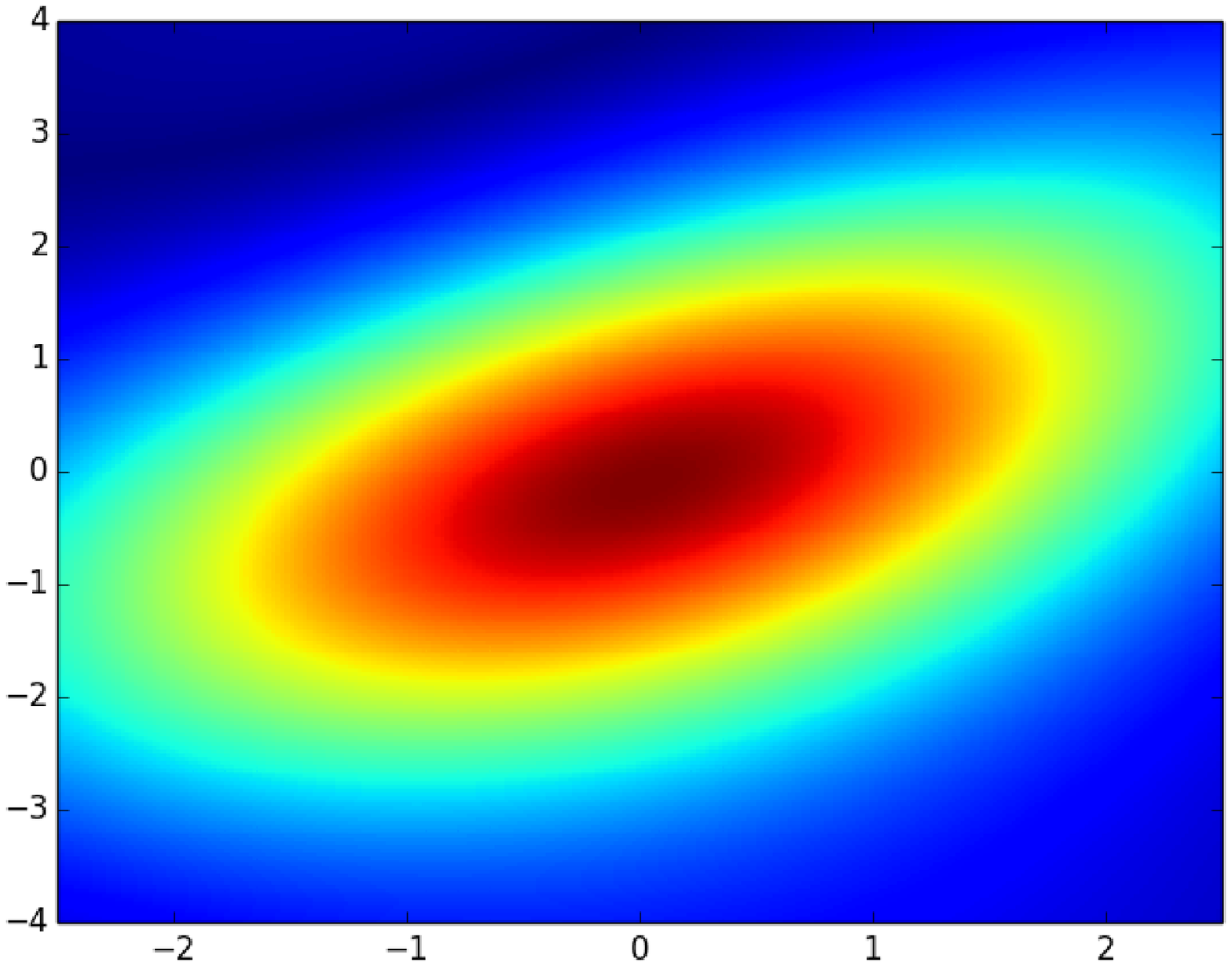}\vskip -3mm
  \onefigure[width=80mm]{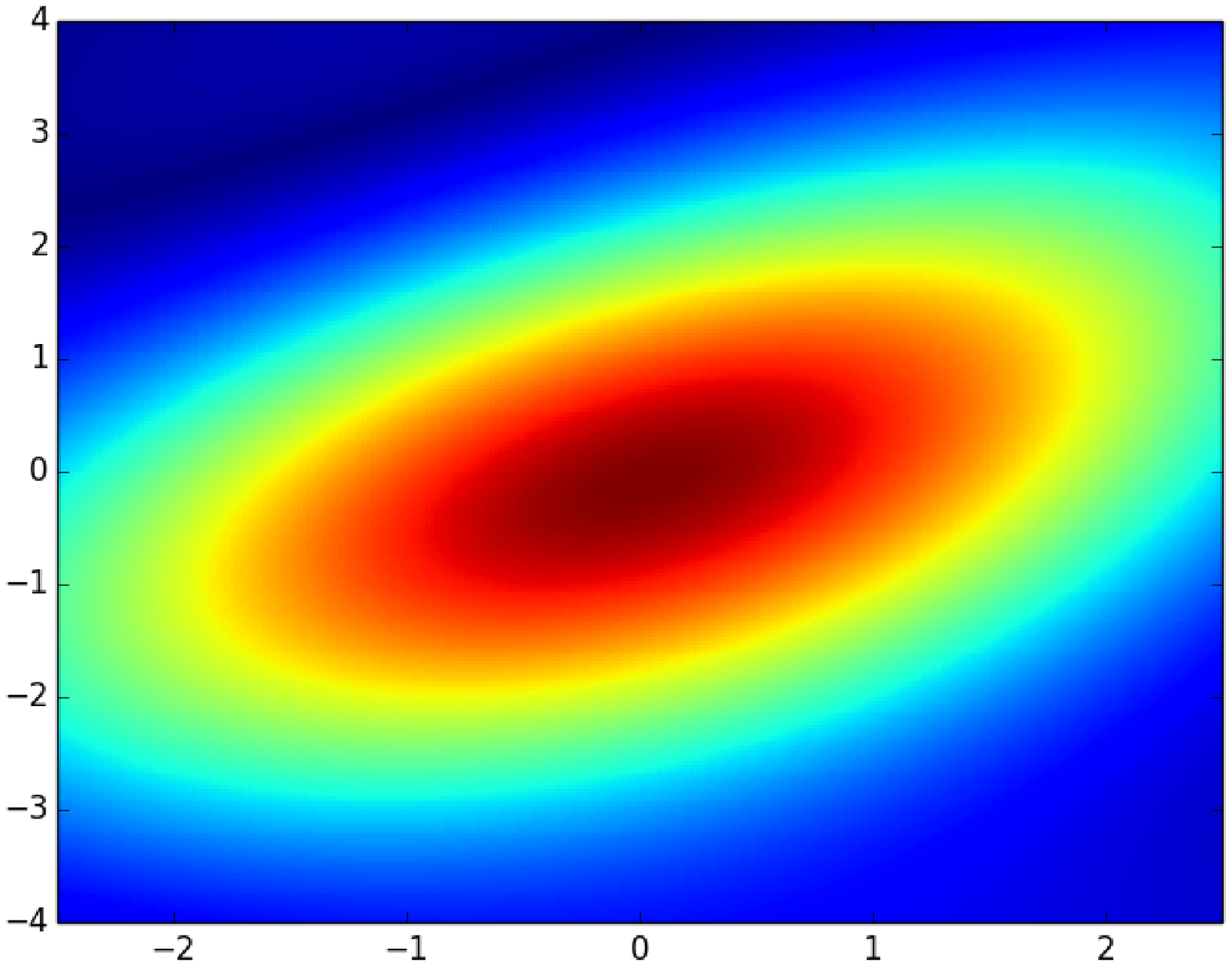}
  \caption{\label{fig:Eigenmodes}Cross sections of the magnetic field amplitude at $z=0$ in the rescaled
    boundary layer coordinates \(\zeta_x, \, \zeta_y \) for $\Rm = 4 \cdot 10^4,
    2\cdot 10^5, 5 \cdot 10^5$.}
\end{figure}

In the asymptotic limit of large \(\Rm\), it is expected that the magnetic
structures will scale as \(\Rm^{-1/2}\) \cite{moffatt1985topological}.
In order to validate this dependency in our direct numerical simulations,
but also to test any additional variation of the leading eigenmode with
\(\Rm\), we produce cross sections through the solution at 
$z=0$ for varying values of \(\Rm\). 
The \textit{loci} of large magnetic field, corresponding to the traces of the
``cigare'' shaped structures on this plane, are then peaks of magnetic
energy. One of these is centred on \( (x=0,\, y=0)\, , \) the section of
this structure in the plane is expected to have a characteristic length-scale
which behaves as \(\Rm^{-1/2}\). The magnetic Reynolds numbers
considered here are extremely large, and the structure is thus sharply localised.
In order to compare the structures
obtained at various values of \(\Rm\), we therefore introduce rescaled coordinates
relevant to the asymptotic limit of large \(\Rm\)
\begin{equation}
\zeta_x = x \, \Rm^{1/2} \, , \qquad
\zeta_y = y \, \Rm^{1/2} \, .
\end{equation}
The magnetic field amplitude is represented versus 
\( (\zeta_x , \, \zeta_y ) \)
for increasing values of the magnetic Reynolds number in
figure~\ref{fig:Eigenmodes}.
\revision{This figure provides instantaneous cross-sections through the cigares, using
rescaled coordinates.}
The leading eigenmode represented in these rescaled coordinates does not
exhibit any significant variation when \(\Rm\) is varied from
\(4\cdot 10^4\) to \(5\cdot 10^5\).
This suggests that the system might be approaching an asymptotic behaviour.
We can however not rule out, on the basis of these numerical simulations, a
remaining slow dependency of the leading eigenmode structure on \(\Rm\) (other than the
length-scale shortening accounted for via the rescaled coordinates).

\section{Damped oscillations}

\begin{figure}
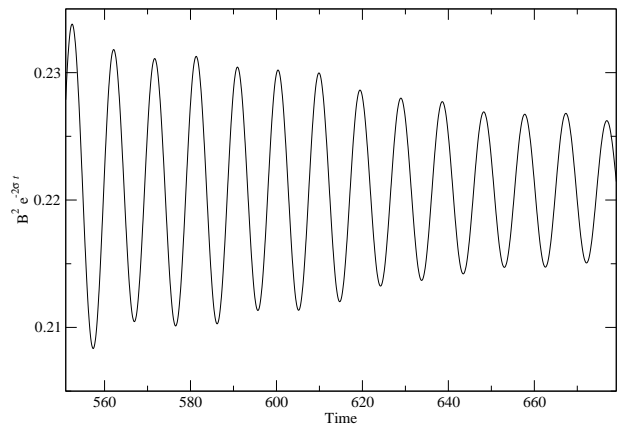

  \onefigure[width=80mm]{200000_osc}
  \caption{\label{fig:osc} Damped oscillations in the time evolution of the magnetic energy corrected
    for the averaged exponential growth rate at \(\Rm=2 \cdot 10 ^5\).}
\end{figure} 

\begin{figure*}
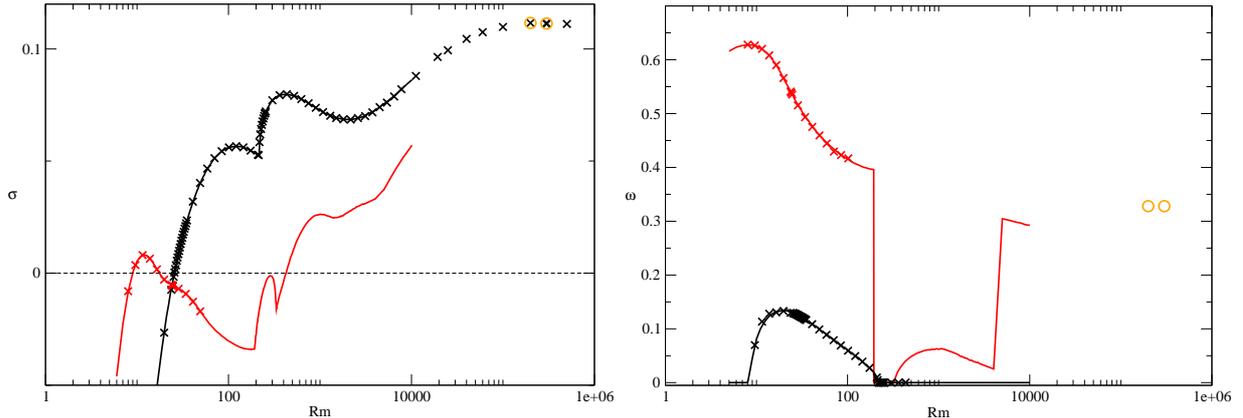

\centerline{\includegraphics*[width=80mm]{Jones_Gilbert_300000_real}
\includegraphics*[width=80mm]{Jones_Gilbert_300000_imag}}
\caption{\label{fig:Jones}Real (left) and imaginary (right) part of the leading eigenvalue
  associated to the V symmetry family (black) and the II symmetry family
  (red), in the classification introduced by \cite{jones2013dynamo}. 
  Continuous lines correspond to the results published in
  \cite{jones2013dynamo}. 
  Crosses present our numerical results in \cite{bouya2013revisiting}
  and in the present study. Circles denote the transient behaviour.
}
\end{figure*} 

A noticeable new feature emerges from the large \(\Rm\) simulations. Whereas the
leading eigenvalue was reported to be purely real for \(\Rm > 215\) (a
critical value denoted \(\Rm_2\) in \cite{bouya2013revisiting}), damped
oscillations appear for \(\Rm > 10^5\) (see figure~\ref{fig:osc}). The
presence of oscillations suggests the existence of a complex eigenvalue. Yet
the fact that these oscillations are damped indicates that the leading
eigenvalue is still real. 

The decomposition in symmetry classes introduced by \cite{jones2013dynamo}
highlighted the families corresponding to the first and the second dynamo
window of the ABC flow, respectively denoted II and V.
Figure~\ref{fig:Jones} suggests that the submode corresponding to this
complex, non-dominant, eigenvalue, could belong to the symmetry class II.

This may be an indication of a new eigenvalue crossing, which could
occur at larger \(\Rm\) and which would result in the reappearance of
time oscillations.  
\revision{Indeed, some models of dynamo action in
  steady flow suggest the possibility of repeated mode crossings as
  \(\Rm \rightarrow \infty\), while the actual growth rate itself
  saturates \cite{finn1988chaotic}.}  
An alternative
scenario, could be that as one is approaching an essential spectrum in
the limit \(\Rm \rightarrow \infty\), \revision{i.e. the complementary to the discrete spectrum
(isolated eigenvalues with finite multiplicity) see \cite{edmunds1986spectral,kato1995perturbation}.}
The growth rate (real part of the
eigenvalue) of all the eigenmodes then tends to the same value.
This
somewhat more optimistic interpretation might suggest that the
asymptotic behaviour of the \(1:1:1\) ABC dynamo, while not yet
obtained numerically could be tackled in a near future.

\section{Discussion}
The values of \(\Rm\) achieved in this study (up to \(5\cdot 10^5\))
are the largest numerically investigated so far. They require a very significant
numerical resolution (up to \(4096 ^3\) Fourier modes) and were performed
using state of the art computational resources.
 
We have first shown that the branch identified in the so called ``second
window'' of the ABC-dynamo (see \cite{galloway1984numerical}) remains the
leading eigenmode up to \(\Rm=5\cdot 
10^5\). This branch corresponds to the V symmetry class introduced by
\cite{jones2013dynamo}, and is associated to a purely real leading
eigenvalue in this parameter range (i.e. \(\Rm \in [215, 5\cdot 10^5]\)).
We show that the growth rate is a decreasing function of \(\Rm\) for the
largest values we could tackle. Furthermore, we demonstrate that the leading eigenmode 
follows the anticipated spatial scaling as \(\Rm^{-1/2}\).
Finally, we indentify slowly damped oscillations occuring at large values of \(\Rm\).

Several aspects of our simulations indicate that the ABC-dynamo is
approaching an asymptotic behaviour for \(\Rm \simeq 10^5\). Namely,
the fact that the cross section through the eigenmode does not reveal
any change in its structure in the rescaled coordinates. This is also
supported by the fact that the growthrate is only slightly above the
theoretical upper bound and is now decreasing with \(\Rm\).  The
occurance of damped oscillations points to an approaching eigenvalue.
This could be relevant to the asymptotic behaviour, for which
an essential spectrum is expected.

However the asymptotic behaviour is not yet established and several issues
indicate that one must be cautious in interpreting the numerical results.
The occurence of slowly damped oscillations, could also be interpreted as a
possible hint for an approaching eigenvalue crossing (as the one observed
near \(\Rm \simeq 24\)). This would result in a change of leading
eigenmode. 
Besides, such high values may still be considered small in some asymptotic problems. Such is the case, for
example in \cite{soward1987fast}, which reveals a decrease of the growth
rate as \( \log(\log(\Rm))/\log(\Rm) \). If this was the case for the
\(1:1:1\) ABC dynamo, its asymptotic behaviour could remain out of reach
of direct numerical simulations for still a long period of time.

Despite its simple analytical form, the ABC-flow dynamo remains remarkably
challenging from a computational point of view. The simulations presented
here are extreme in terms of parameter value (fast dynamo limit), in terms of
numerical resolutions (up to $4096^3$) and in terms of computational
resources (more than $10^6$ CPU hours for the five new values of \(\Rm\)
calculated in this study).

The two main scenario that emerge from our study, are either the
possibility of an approaching new mode crossing which implies that the
asymptotic behaviour is not yet met, or that the real parts of all eigenvalues are
approaching the same limit, which would instead indicate
an essential spectrum.
Simulations at larger values of \(\Rm\) may shed some light on those issues.

\acknowledgements
The authors want to acknowledge the access 
the {\it MesoPSL Challenge} which allowed the present study.
\revision{The HPC resources of MesoPSL were financed
by the Region Ile de France and the project Equip@Meso (reference
ANR-10-EQPX-29-01) of the programme Investissements d'Avenir supervised
by the Agence Nationale pour la Recherche. }

The authors are grateful to Prof. Andrew Gilbert for useful discussions.

\bibliographystyle{alpha}
\bibliography{biblio_perso}

\begin{thebibliography}{ADN03}

\bibitem[ADN03]{archontis2003numerical}
Vasilis Archontis, S{\o}ren Bertil~Fabricius Dorch, and {\r A}ke Nordlund.
\newblock Numerical simulations of kinematic dynamo action.
\newblock {\em Astronomy and Astrophysics}, 397(2):393--399, 2003.

\bibitem[Arn65]{arnold1965topologie}
Vladimir~Igorevich Arnol'd.
\newblock Sur la topologie des {\'e}coulements stationnaires des fluides
  parfaits.
\newblock {\em Comptes rendus hebdomadaires des s{\'e}ances de l'Acad{\'e}mie
  des Sciences}, 261(1):17--20, July 1965.

\bibitem[BD13]{bouya2013revisiting}
Isma{\"e}l Bouya and Emmanuel Dormy.
\newblock Revisiting the {ABC} flow dynamo.
\newblock {\em Physics of Fluids}, 25(3), 2013.

\bibitem[CG95]{childress1995stretch}
Stephen Childress and Andrew~D. Gilbert.
\newblock {\em Stretch, twist, fold: the fast dynamo}, volume~37 of {\em
  Lecture Notes in Physics Monographs}.
\newblock Springer Berlin Heidelberg, 1995.

\bibitem[EE86]{edmunds1986spectral}
David~E. Edmunds and Will~D. Evans.
\newblock {\em Spectral theory and differential operators}.
\newblock Oxford University Press, 1986.

\bibitem[FO88]{finn1988chaotic}
John~M. Finn and Edward Ott.
\newblock Chaotic flows and fast magnetic dynamos.
\newblock {\em Physics of Fluids}, 31(10):2992, 1988.

\bibitem[GF84]{galloway1984numerical}
David~J. Galloway and Uriel Frisch.
\newblock A numerical investigation of magnetic field generation in a flow with
  chaotic streamlines.
\newblock {\em Geophysical \& Astrophysical Fluid Dynamics}, 29(1-4):13--18,
  1984.

\bibitem[H{\'e}n66]{henon1966topologie}
Michel H{\'e}non.
\newblock Sur la topologie des lignes de courant dans un cas particulier.
\newblock {\em Comptes rendus hebdomadaires des s{\'e}ances de l'Acad{\'e}mie
  des Sciences}, 262:312--314, January 1966.

\bibitem[JG14]{jones2013dynamo}
Samuel~E. Jones and Andrew~D. Gilbert.
\newblock Dynamo action in the {ABC} flows using symmetries.
\newblock {\em Geophysical \& Astrophysical Fluid Dynamics}, 108(1):83--116,
  2014.

\bibitem[Kat95]{kato1995perturbation}
Tosio Kat{\=o}.
\newblock {\em Perturbation theory for linear operators}.
\newblock Classics in Mathematics. Springer, 1995.

\bibitem[KY95]{klapper1995rigorous}
Isaac Klapper and Lai-Sang Young.
\newblock Rigorous bounds on the fast dynamo growth rate involving topological
  entropy.
\newblock {\em Communications in Mathematical Physics}, 173(3):623--646, 1995.

\bibitem[MP85]{moffatt1985topological}
Henry~K. Moffatt and Michael R.~E. Proctor.
\newblock Topological constraints associated with fast dynamo action.
\newblock {\em Journal of Fluid Mechanics}, 154:493--507, 1985.

\bibitem[Sow87]{soward1987fast}
Andrew~M. Soward.
\newblock Fast dynamo action in a steady flow.
\newblock {\em Journal of Fluid Mechanics}, 180:267--295, 1987.

\end{thebibliography}

\end{document}